\documentclass[prb,twocolumn]{revtex4} 

\usepackage[latin1]{inputenc}
\usepackage[T1]{fontenc}
\usepackage[english]{babel}
\usepackage{times}
\usepackage{graphicx}

\def\CoIII{Co\textsuperscript{3+}}
\def\CoIV{Co\textsuperscript{4+}}

\def\CoLIX{{\textsuperscript{59}Co}}
\def\NaxCoOO{Na$_x$CoO$_2$}

\def\ETAL{{\it et al.}}

\begin{document}
\title{Spin correlations and cobalt charge states: A new phase diagram of sodium cobaltates}
\author{G. Lang,\textsuperscript{1,*} J. Bobroff,\textsuperscript{1} H. Alloul,\textsuperscript{1} G. Collin,\textsuperscript{2} and N. Blanchard\textsuperscript{1}}
\affiliation{
\textsuperscript{1}Laboratoire de Physique des Solides, UMR 8502, Université Paris-Sud, 91405 Orsay, France \\
\textsuperscript{2}Laboratoire Léon Brillouin, CE Saclay, CEA-CNRS, 91191 Gif-sur-Yvette, France
}

\begin{abstract}
Using {$^{23}$Na} NMR measurements on sodium cobaltates at intermediate dopings (0.44$\le$$x$$\le$0.62), we establish the qualitative change of behavior of the local magnetic susceptibility at $x^*$=0.63--0.65, from a low $x$ Pauli-like regime to the high $x$ Curie-Weiss regime.
For 0.5$\le$$x$$\le$0.62, the presence of a maximum $T^*$ in the temperature dependence of the susceptibility shows the existence of an $x$-dependent energy scale.
$T_1$ relaxation measurements establish the predominantly antiferromagnetic character of spin correlations for $x$$<$$x^*$.
This contradicts the commonly assumed uncorrelated Pauli behavior in this $x$ range and is at odds with the observed ferromagnetic correlations for $x$$>$$x^*$.
It is suggested that at a given $x$ the ferromagnetic correlations might dominate the  antiferromagnetic ones above $T^*$.
From {$^{59}$Co} NMR data, it is shown that moving towards higher $x$ away from $x$=0.5 results in the progressive appearance of nonmagnetic \CoIII\ sites, breaking the homogeneity of Co states encountered for $x$$\le$0.5.
The main features of the NMR-detected {$^{59}$Co} quadrupolar effects, together with indications from the powder x-ray diffraction data, lead us to sketch a possible structural origin for the \CoIII\ sites.
In light of this ensemble of new experimental observations, a new phase diagram is proposed, taking into account the systematic presence of correlations and their $x$-dependence.
\end{abstract}

\maketitle

\section{Introduction}

After having been extensively studied as materials for batteries,\cite{Fouassier1973} sodium cobaltates attracted attention due to their high thermoelectric power\cite{Terasaki1997} together with rather good metallicity. However, with hints of strong correlations from specific heat measurements\cite{Ando1999} and with the discovery of a superconducting phase,\cite{Takada2003} a surge in interest has occured in the past few years.
A major reason to this is that \NaxCoOO\ phases are layered transition metal oxides of a $3d$ metal, thus reminding one of the high-$T_c$ cuprates.
Indeed, they are made of a stacking of oxygen and cobalt layers, forming CoO$_2$ layers of edge-connected CoO$_6$ octahedra. These layers are themselves separated by sodium layers, in which two different sites can generally be occupied with respect to the nearest cobalt layers.
As the cobalt layers feature a triangular lattice, effects linked with the frustration of magnetic interactions between cobalts might be expected.
By varying the sodium content, the doping of the cobalt layers can be modified, yielding a rich phase diagram.
In the $x$=1 limit, the material is a band insulator with only nonmagnetic $S$=0 \CoIII\ ions.\cite{Lang2005}
As sodium is removed, a Curie-Weiss susceptibility behavior appears\cite{Carretta2004,Ihara2004} down to supposedly $x$=0.67, together with long-range ordered magnetic states ($T_N$=19--27~K)\cite{Sugiyama2003a,Bayrakci2004,Sugiyama2004,Bayrakci2005,Mendels2005} for $x$=0.75--0.9.
In the $x$$\approx$0.7 region, intralayer correlations of spin fluctuations have been shown to be those of a two-dimensionnal quasi-ferromagnetic metal.\cite{Alloul2008}
On contrary to the case of cuprates, but as in that of manganites, doping in this $x$ region results in differently charged Co sites \cite{Mukhamedshin2005}. The corresponding cobalt charge ordering is presumably linked to the structural Na order.
Going to lower $x$, a cross-over to Pauli-like susceptibility and to homogeneous doping takes place.
It is suggested\cite{Foo2004} to occur at $x$=0.5, which features\cite{Mendels2005} additionaly magnetic ($T_N$=86~K) and metal-insulator ($T_{MIT}$=51~K) transitions.
This susceptibility behavior extends to $x$=0.35, the precursor of the hydrated superconductor, and possibly down to the Mott insulator limit $x$=0.\cite{Vaulx2007}

Adding sodium to CoO$_2$ would naively be expected to increase the proportion of nonmagnetic cobalts by turning \CoIV\ ions into \CoIII\ ions.
Therefore, a surprising property of the phase diagram is the apparent increase of magnetic correlations (Curie-Weiss behavior and frozen magnetic states) with increasing $x$.
Concurrently, the above simplistic \CoIV/\CoIII\ ionic description appears experimentally valid only in the $x=1$ limit, for which all cobalts are \CoIII.\cite{Lang2005}
In this regard, it appears highly desirable to scrutinize whether $x=0.5$ is a threshold both for the susceptibility behavior and for the occurrence of inhomogeneous Co states.
Especially, as the specific Na order has been suggested to drive or contribute to the magnetic and metal-insulator transitions through Fermi surface reconstruction,\cite{Bobroff2006,Balicas2005,Zhou2007} the $x$=0.5 phase may be a pathologic case, irrelevant to the physics on either side of the phase diagram.
More recent work has suggested\cite{Yokoi2005,Yoshizumi2007,Shu2007} that the boundary between Pauli and Curie-Weiss behavior is closer to $x^*$$\approx$0.6.
Unfortunately, as this was done using only macroscopic measurements (susceptibility, specific heat), it cannot be ruled out that there is some mixture of phases with different Na contents or order, should the latter be of importance in this respect.
This could naturally smooth the transition between the two susceptibility behaviors, with an uncertain boundary.
Local static and dynamic susceptibility measurements are thus required to determine the actual value of $x^*$ and to establish whether the onset of Co charge
differentiation coincides or not with the crossover in correlation regime.
Combined with structural characterization, one might then determine whether the corresponding charge ordering is linked to sodium ordering.
Furthermore, the Pauli-like or noncorrelated nature of the phases with $x$$<$0.5 appears doubtful.
Indeed, while the Mott insulator limit for $x=0$ has not yet been evidenced, strong AF correlations have been found\cite{Vaulx2007} for very low Na content ($x\rightarrow 0$) with an evolution towards a renormalized Fermi liquid below $T^*$$\approx$7~K.
Establishing whether the difference in static spin susceptibility for low and high sodium content coincides with a difference in spin correlations may then yield insight on the physics at play.

To address these questions to some extent, we present in this paper nuclear magnetic resonance (NMR) results obtained on sodium cobaltates in the intermediate range of dopings $0.44 \le x \le 0.62$, which bridges the two important domains of the phase diagram.
Let us emphasize that only limited investigation of this doping range has been performed so far. Furthermore the present study is the first one to probe the local properties, and will enable us to suggest a new phase diagram for the Na cobaltates from the behaviors of both the static and dynamic susceptibilities.
In section \ref{sec:structure}, we briefly give details on sample preparation and present x-ray diffraction measurements supporting an incommensurate modulation of the Na-layer order.
Using sodium NMR, in section \ref{sec:chi_spin}, we establish locally that there exists a well-defined boundary at $x^*$$\approx$0.6 between the Pauli-like and Curie-Weiss behavior.
Furthermore, it is shown that there exists an $x$-dependent energy scale, at least for 0.5$\le$$x$$\le$0.62, characterized by the temperature $T^*$ for which the spin
susceptibility goes through a maximum.
In section \ref{sec:correl}, measurement and analysis of the temperature dependence of the sodium $T_1$ NMR relaxation time allows us to demonstrate the predominantly antiferromagnetic character of spin correlations for $x$$<$$x^*$.
In section \ref{sec:Co_states}, cobalt NMR is used to probe directly the cobalt charge and spin states, showing the progressive apparition of nonmagnetic \CoIII\ sites for sodium content above $x$=0.5.
Putting this result in perspective with crystallographic characterization, we suggest a simple frame in which to describe the local structure.
Finally, these results are summarized and discussed in sections \ref{sec:diagram} and \ref{sec:discussion}, where a new phase diagram summing up the main physics problems in sodium cobaltates is proposed.

\section{Crystallographic characterization}
\label{sec:structure}

All studied samples were powders prepared via a solid state reaction.
For $x$=0.5, both the orthorhombic ($\gamma$ phase) and the rhombohedral ($\beta$ phase) structures were used.
These will be referred to respectively as ``O'' and ``R'' on figures.
The $\gamma$ phase was prepared as described in a previous paper,\cite{Mendels2005} while the rhombohedral phase was prepared by sodium deintercalation, in a concentrated sodium hypochlorite solution, of the rhombohedral $x$=1 phase. The synthesis of the latter has also been previously described.\cite{Lang2005}
All other compositions, namely $x$=0.44, 0.55, 0.58 and 0.62, were of the hexagonal type, i.e. the same structural family as the $x$=0.5 $\gamma$ phase.
Those were obtained by annealing of a mixture of this $\gamma$ phase with $x$=0.71 material, under O$_2$ flux and at $T$=180--350ï¿½C, thus several-week long reaction times.
For NMR purposes, sample cristallites were aligned along their $c$ axis in Stycast 1266 epoxy.
We should immediately emphasize that this is the first study of such intermediate dopings in which the single-phase character of samples has been verified not only by x-ray diffraction but also locally using $\mu$SR and NMR, in keeping with previous experiments performed in our group.\cite{Mendels2005,Alloul2008}
Providing feedback to the x-ray characterization, this allowed us for instance to avert illegimate datapoints for initial compositions close to $x$=0.53, which were shown to be containing a mixture of the magnetically-frozen $x$=0.5 phase and of nonmagnetic phases with larger Na content.

\begin{figure}[pbth]
\includegraphics[width=8.5cm]{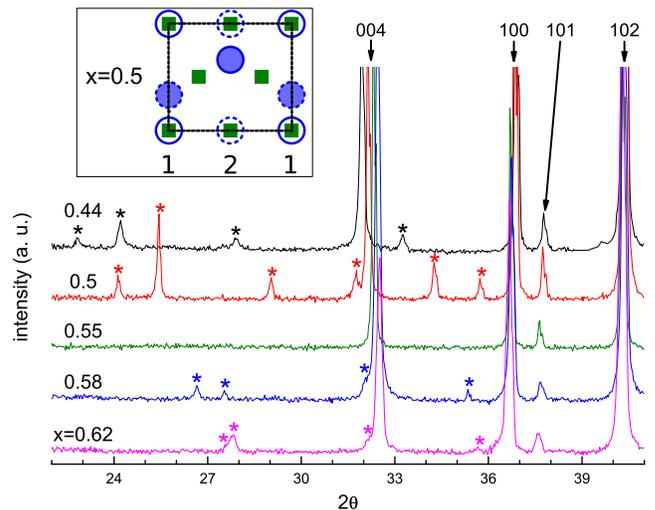}
\caption{(Color online) x-ray powder diffraction diagram for the studied hexagonal phases. $hkl$ indices are indicated for the Bragg peaks corresponding to the mean structure, while stars designate satellites linked to the modulation of the Na layers. The unit cell of the $x$=0.5 phase is shown in the insert. Cobalts are marked as green squares, sodiums as blue circles (an open/full circle indicates a Na1/Na2 site, with a full/dashed line indicating this Na is above/below the cobalts). The bottom labels refer to the Na1 and Na2 bands in the upper plane.}
\label{fig:RX_05x}
\end{figure}

\begin{figure}[pbth]
\includegraphics[width=7.5cm]{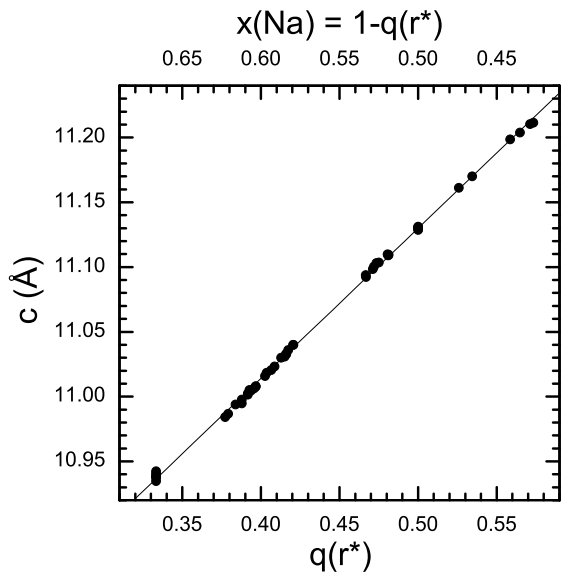}
\caption{Relation between the $c$ lattice parameter and the magnitude of the modulation wavevector $q(r^*)$. The top axis yields $x$ through the relation $x$=1-$q(r^*)$. The solid line is a linear fit of the full dataset.}
\label{fig:c_x_modul}
\end{figure}

Structural characterization was performed by x-ray diffraction at room temperature using the Cu K-$\alpha$ radiation.
Part of the powder diagrams is shown on Figure \ref{fig:RX_05x}.
They feature Bragg peaks corresponding to the reference hexagonal structure (the mean structure), together with additional ensembles of reflections (stars on Fig.~\ref{fig:RX_05x}) except for $x$=0.55.
These extra peaks demonstrate the existence of a commensurate superstructure in the Na layers for $x$=0.5, identical to the already well-known structure (see inset of Fig.~\ref{fig:RX_05x}).\cite{Huang2004b,Zandbergen2004}
In the $x$$>$0.55 phases, the superstructure is on the contrary incommensurate.
Through refinement of the powder diagrams, one can extract the magnitude of the corresponding modulation wavevectors $q(r^*)$ in addition to the lattice parameters.
Similar observations were made for $x$$\approx$0.7, although the details of the modulation tend to get more complicated for these larger $x$ values.\cite{Collin200x}
We found that a linear relation holds between $q(r^*)$ and the $c$ lattice parameters measured for all studied samples, as can be seen in Fig.~\ref{fig:c_x_modul} (see also Tab.~\ref{tab:parameters} for some values).
For the better known phases such as $x$=0.5, $x$=0.67 and $x$=0.72, the $c$ lattice parameters measured on samples studied by our group are in good agreement with values from the literature.
Using their $x$ values one then finds quite remarkably that $x$=1-$q(r^*)$.
The linear relation between $q(r^*)$ and $c$ can then be rewritten likewise (top axis of Fig.~\ref{fig:c_x_modul}), reflecting the already known fairly linear relation between $x$ and $c$.
Accordingly, the $x$ values deduced from $q(r^*)$ are found to coincide with the proportions of materials used in the initial syntheses.
Compared to the usual $c$ axis measurement, such a relation allows for much better $x$ determination at intermediate dopings, since for a given $x$ variation the relative variation of $q(r^*)$ is much larger than that of $c$.

\begin{table}
\begin{center}
\begin{tabular}{cccc}
\hline \hline
$x$ & $a$ (\AA) & $c$ (\AA) & $q(r^*)$ \\
\hline
0.44 & 2.81298(4) & 11.2039(2) & 0.5648(2) \\
0.50 & 2.81519(3) & 11.1308(2) & 0.5000(0) \\
0.55 & 2.81940(3) & 11.0676(2) & - \\
0.58 & 2.82171(2) & 11.0400(1) & 0.4206(1) \\
0.62 & 2.82468(5) & 10.9940(3) & 0.3839(3) \\
\hline \hline
\end{tabular}
\caption{Lattice parameters of the mean hexagonal structure and value of the modulation wave vector for the studied hexagonal samples. $x$ is directly extracted from $q(r^*)$.}
\label{tab:parameters}
\end{center}
\end{table}

While the precise details of the modulation are still not known with certainty, it appears clearly from the initial analysis that it is related to a periodic stripe organization of the Na layers.
In this picture, the $x$=0.5 phase is a central point with an equal number of Na1 and Na2 bands (see inset of Fig.~\ref{fig:RX_05x}).
On increasing $x$ above 0.5, additional bands of Na2 ions appear, while below 0.5 additional bands of vacancies show up.
The fact that the Na superstructures can be detected, except for $x$=0.55, by x-ray diffraction implies that the long-range correlations of the modulation extend in all three directions at room temperature, i.e. not only in the Na layers but also along $c$ with spatial correlations of the extra Na2 bands.
We will show using Co NMR that comparable correlations are also present for $x$=0.55 at low temperature.
Additionaly, we will discuss possible implications of such a modulation on Co charge states.

\begin{figure}[pbth]
\includegraphics[width=8.5cm]{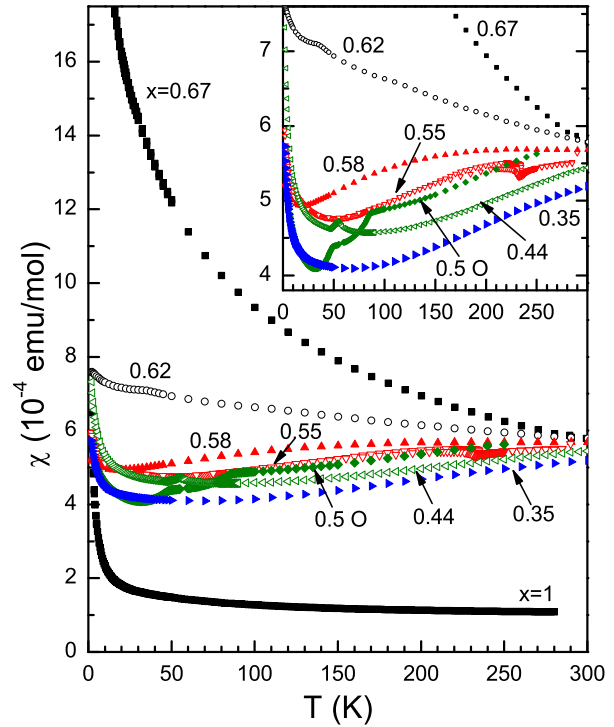}
\caption{(Color online) Temperature dependence of the macroscopic magnetic susceptibility, as measured with field-cooling and an applied field of 10~kG, save for $x$=1 (1~kG) and $x$=0.58/0.62 (50~kG). The insert shows on an expanded scale the progressive evolution at intermediate dopings ($x$=1 has been omitted for low-$T$ clarity). Singularities due to intrinsic ordering ($x$=0.5) and spurious phases are better seen here. The $x$=0.5, $x$=0.67 and $x$=1 data are from (or similar to those of) previous publications.\cite{Mukhamedshin2004,Lang2005,Bobroff2006}}
\label{fig:squid}
\end{figure}

\section{Doping-dependence of spin susceptibility}
\label{sec:chi_spin}

A first look at the $x$-dependence of the magnetic suspectibility can be given by SQUID (Superconducting QUantum Interference Device) measurements. On Figure \ref{fig:squid}, the temperature dependence for the studied hexagonal phases is shown with the Pauli-like $x$$=$0.35, the Curie-Weiss $x$$=$0.67 and the nonmagnetic $x$$=$1 phases for reference.
Starting from $x$=0.35, increasing the sodium content results in a continuous evolution towards the Curie-Weiss behavior, with a clear acceleration at $x^*$$\approx$0.6, in qualitative agreement with previous determinations.\cite{Yokoi2005,Yoshizumi2007,Shu2007}
We do not of course take into account the temperature domain where magnetic freezing takes place at $x$=0.5, i.e. below $T$=86~K.
However, even in well-characterized samples one may not entirely rule out a small phase mixture, either with another cobaltate phase as mentioned in the introduction, or with some impurity phase.
For instance, it can be seen that small kinks occur around $T$=55~K for $x$=0.44, possibly due to frozen O$_2$, and around $T$=30~K for $x$=0.62, possibly due to Co$_3$O$_4$ traces.
On the contrary, no such indications are present for $x$=0.55 and $x$=0.58.
Note also that we cannot, from macroscopic measurements, ascertain the extrinsic character of the low-temperature Curie tails, which are likely due to spurious phases as in the $x$=1 phase.
Furthermore, even for a single phase, the measured susceptibility accounts not only for the spin susceptibility (essentially from the cobalt planes), but also for a fairly $x$-independent and predictable diamagnetic contribution of the ion cores, and a Van-Vleck type orbital contribution:
$$
\chi = \chi_{spin} + \chi_{dia} + \chi_{VV}
$$
The latter, while in principle temperature-independent, is hard to estimate which does not allow us to perform a fully reliable estimate of the temperature dependence of the spin contribution $\chi_{spin}$.

\begin{figure}[pbth]
\includegraphics[width=8.5cm]{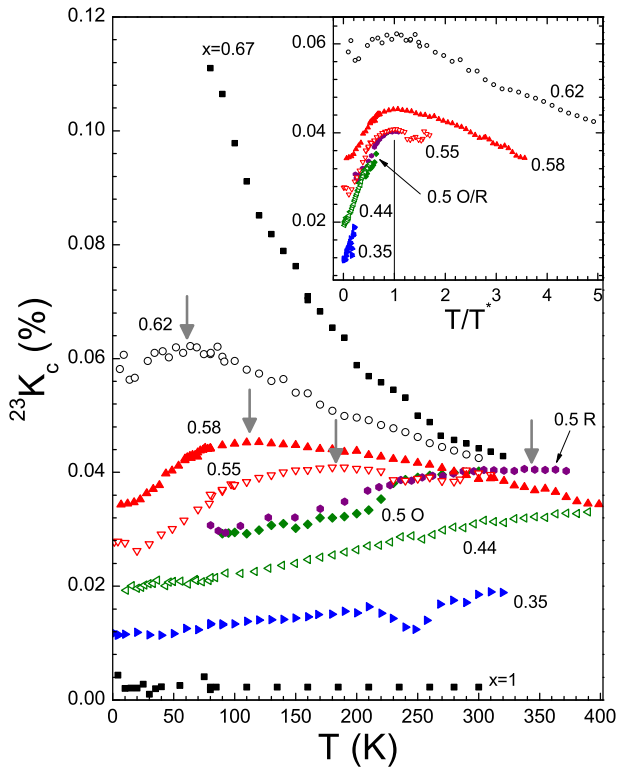}
\caption{(Color online) Temperature dependence of the sodium NMR shift. Down-pointing arrows indicate the maxima for $x$=0.5--0.62 (for $x$=0.5, data at high $T$ has only been taken on the rhombohedral (R) structure sample). The corresponding temperatures $T^*$ are used for rescaling the data, as shown in the inset. The $x$=0.35, $x$=0.5 O, $x$=0.67 and $x$=1 data are from (or similar to those of) previous publications.\cite{Mukhamedshin2004,Lang2005,Bobroff2006}}
\label{fig:KNa}
\end{figure}

To lift these ambiguities, sodium NMR is an appropriate tool since it allows one to probe the spin susceptibility without bias from the Van-Vleck contribution, while helping to ensure the single-phase character of samples through individual site detection.
Likewise, the extrinsic character of Curie tails can be verified.
Indeed, thanks to numerous hyperfine couplings through Na--O--Co superexchange paths,\cite{Johannes2005,Ning2005,Kroll2006} the sodium nuclei are directly probing the average intrinsic cobalt planes spin susceptibility $\chi_{spin}$, as was clearly shown in previous NMR studies.\cite{Carretta2004,Mukhamedshin2004,Mukhamedshin2005}
In practice, this is measured through the shift ${^{23}K}$ of the sodium nuclear resonance line:
\[
{^{23}K} = \frac{{^{23}A_{hf}} \cdot \chi_{spin}}{{\cal N}_A \cdot \mu_B}
\]
with $\chi_{spin}$ in emu/mol, ${^{23}A_{hf}}$ the hyperfine coupling constant, ${\cal N}_A$ the Avogadro number, and $\mu_B$ the Bohr magneton.
Experimentally, NMR spectra were measured on fixed-frequency and fixed-field setups, using $\frac{\pi}{2} - \tau - \frac{\pi}{2}$ spin echo sequences. Fourier tranformation was applied to each individual echo, allowing for precise shift determination.
As frozen magnetism occurs at low $T$ for $x=0.5$, we searched for such a possibility for $x$$\ne$0.5 using Na and Co NMR measurements as well as $\mu$SR. Together with a suppression of the paramagnetic volume fraction seen by $\mu$SR, we could then expect, for instance, a splitting of the NMR spectra as seen for $x$=0.5.
However, no evidence of frozen magnetism was detected for any of the $x$$\ne$0.5 phases down to at least $T$=10~K.
The temperature-dependence of the sodium shift along the $c$ direction (see main panel of Figure \ref{fig:KNa}), as measured on the central line, yields similar results to that of SQUID measurements, lifting any uncertainty: there is indeed a clear boundary at $x^*$$\approx$0.6 separating two fairly-well defined domains of magnetic correlations.
It is worth noting that the low sodium content domain ($x$$<$$x^*$) is clearly not strictly Pauli, suggesting for instance significant correlation effects.
$x^*$ would then correspond to a change in the correlation regime, rather than the mere appearance of correlations at high sodium content.
A new and unexpected information is the presence for 0.55$\le$$x$$\le$0.62 of a temperature $T^*$ for which the susceptibility goes through a maximum, as indicated by thick down-pointing arrows, with $T^*$ decreasing for increasing $x$.
High-temperature data for the rhombohedral $x$=0.5 phase seems to indicate a similar behavior around $T$$=$340~K.
We cannot exclude a comparable situation for $x$=0.35 and $x$=0.44, however an experimental verification would require temperatures significantly beyond room-temperature, which would likely alter the samples.
As is shown in the inset of Fig.~\ref{fig:KNa} where the temperature axis has been renormalized\footnote{Somewhat speculatively, the $x$=0.35 and $x$=0.44 phases have been included, with $T^*$ adjusted for slope matching at low $T/T^*$. The corresponding $T^*$ values are 1500(600)~K and 750(300)~K, whose physical relevance is unclear.}
 by $1/T^*$, there is a clear similarity of behavior for all dopings.
It thus appears that the $x$$\le$$x^*$ region is characterized by an $x$-dependent energy scale, defined by $T^*$ which tends toward 0 as $x$ approaches $x^*$ (the doping dependence of $T^*$ will be presented in Sec.~\ref{sec:diagram}).
The existence of a susceptibility maximum is indicative of a threshold in the correlations of these systems.
While for $T$$\gg$$T^*$ one may wonder whether the high $x$ Curie-Weiss behavior is restored, at $T$$<$$T^*$ there is a sizable decrease of $\chi_{spin}$ which is indicative of a progressive freezing of low-energy excitations.

\section{Nature of spin correlations}
\label{sec:correl}

Having shown the clear separation into two correlation regimes at the boundary $x^*$$\approx$0.6, it is desirable to better differentiate the two regions.
This can be done through measurement of the Na nuclear spin lattice relaxation time $T_1$.
Indeed, just as its NMR shift is a good probe of the cobalt uniform ($\mathbf q$=0) spin susceptibility, sodium reflects through $T_1$ the local field fluctuations associated with the electronic dynamic spin susceptibility, i.e. at $\mathbf q$$\ne$0.
While it is known from such Na NMR relaxation measurements\cite{Alloul2008} that the 0.67$\le$$x$$\le$0.75 phases show two-dimensionnal quasi-ferromagnetic behavior in the paramagnetic state, this may no longer apply for $x$$<$$x^*$.
Besides, such a study needs to take into account a possible additional contribution to the local field fluctuations, linked to Na ionic diffusion, as was shown to be the case for temperatures above 200~K in the 0.67$\le$$x$$\le$0.75 phases.
In that case we can write:\cite{Abragam1961}
\[
\frac{1}{T_1} = \left(\frac{1}{T_1}\right)_e + \left(\frac{1}{T_1}\right)_d
\]
where $e$ indicates the relaxation due to conduction electrons, proportional to $T$ in a usual metal, and $d$ the relaxation due to Na diffusion.
Therefore, for 0.44$\le$$x$$\le$0.62, we will show in a first subsection the occurrence of Na motion and its extent, which will allow us to analyze in a second subsection the electronic contribution to $T_1$.

\subsection{Na motion}

\begin{figure}[pbth]
\includegraphics[width=8cm]{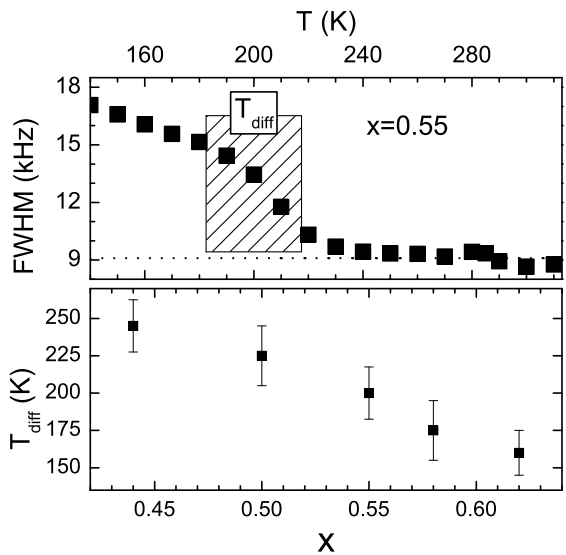}
\caption{(upper panel) Na ionic diffusion as seen from the Na NMR central line:  typical temperature dependence of the linewidth, here for the $x$=0.55 main Na site along the $c$ direction, with the dotted line indicating the intrinsic linewidth. (lower panel) Doping dependence of the corresponding $T_{diff}$ diffusion temperature.}
\label{fig:Na_diffusion}
\end{figure}

The existence of Na diffusion processes can easily be seen from the sharpening of the Na central line in the NMR spectra, due to motional narrowing at the right frequency, as can be seen for instance for $x$=0.55 in the upper panel of Fig.~\ref{fig:Na_diffusion}.
This narrowing occurs here at $T_{diff}$$\approx$180--220~K, and depends on the sodium content (see the lower panel of Fig.~\ref{fig:Na_diffusion}).
Note that the decrease of $T_{diff}$ with increasing sodium content appears counter-intuitive, and may suggest an increasingly unstable Na order as the incommensurate modulation vector shortens.

Let us now study the contribution of these diffusion processes to the Na spin lattice relaxation time. The $T_1$ measurements were performed  using the saturation recovery technique for the central transition (- 1/2, 1/2) of the NMR spectrum.
Assuming a purely magnetic relaxation, we would expect a magnetization relaxation profile following:\cite{McDowell1995,Roscher1996}
\[
M(t) = M_0 \cdot \left[ 1 - B\left((1-W) e^{-t/T_1} + W e^{-6t/T_1}\right) \right]
\]
with $W$=0.9 for a $\pi/2$ saturation pulse and $B$ a correction factor for imperfect saturation.
A single $\pi/2$ saturation pulse was found to be sufficient, with typically 0.9$<$$B$$<$1.1.
We observed significant variations of $W$ (typically 0.5$\le$$W$$\le$0.85), even for a single sample.
In the latter case this was very likely due to the additional contribution from sodium ionic diffusion at high temperature, which weakens the above hypothesis of a purely magnetic relaxation.
Corresponding high temperature $T_1$ values should then be taken with caution for quantitative analysis.
For the sake of comparison we forced $W$$\approx$0.7, with the error bars reflecting the drift in $W$ values, thus to be read as a confidence interval.

\begin{figure}[pbth]
\includegraphics[width=8cm]{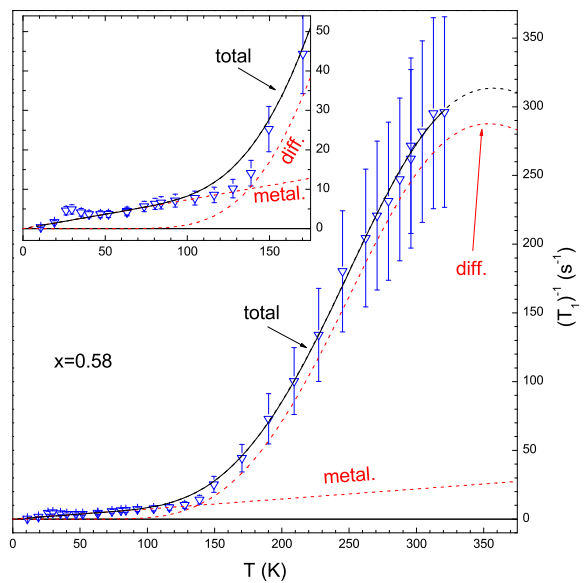}
\caption{(Color online) Temperature dependence of the sodium $T_1^{-1}$ relaxation rate for $x$=0.58, along the $c$ direction. Drawn as a black line is a fit (see text) which is the sum of the sodium diffusion contribution (``diff.'') and the metallic contribution (``metal.''), both drawn as red lines. The inset shows the negligible contribution of ionic diffusion at low temperature.}
\label{fig:T1_058}
\end{figure}

On Fig.~\ref{fig:T1_058}, we show the temperature dependence of the relaxation rate $T_1^{-1}$ for $x$=0.58. The strong increase of the relaxation above $T$$\approx$100~K, which cannot be ascribed to electronic processes as nothing so spectacular is observed for the NMR shift, is likely associated with the planar diffusion of Na ions.
Similar observations could be made for the other studied phases.
The $x$=0.58 data was thus fitted using a linear term (metallic contribution, see above) and a Bloembergen-Purcell-Pound like diffusion term, as suggested by de Vaulx {\it et al.}\cite{Vaulx2007a} and used for instance in the study of Li$_x$CoO$_2$:\cite{Nakamura2006}
\[
\left(\frac{1}{T_1}\right)_d = C \left[ \frac{\tau}{1+(\omega \tau)^n} + \frac{4 \tau}{1+(2 \omega \tau)^n} \right]
\]
where $C$ is a coupling constant and $\tau$ is the characteristic diffusion time, assumed to be $\tau$=$\tau_0 \cdot \exp(U/k_BT)$ with $U$ the activation energy.
The total fit (``total'' in Fig.~\ref{fig:T1_058}) shows that the metallic contribution (``metal.''), while important at low temperature, is indeed largely superseded by the diffusion term (``diff.'') at high temperature.
We find $C$=10$^{11}$ (fixed), $\tau_0$=2$\:$10$^{-11}$~s, U$\approx$140~meV and n=1.54, in reasonable agreement with $\tau_0$=6.2$\:$10$^{-10}$~s, $U$=80~meV and $n$=1.34 as measured in Li$_{0.8}$CoO$_2$.\cite{Nakamura2006}
As noted above this is an approximate procedure, since the high-temperature $T_1$ values are then derived from partly inappropriate fits of the $M(t)$ magnetization profiles.
We do not intend here to study in more detail the Na diffusion processes, which are highly dependent of the Na content.
It remains however that this preliminary study shows that it is safe to expect no significant diffusion contribution to 1/$T_1$ at low temperature (below $T$$=$50--150~K depending on the sample).

\subsection{Electronic processes}

\begin{figure}[pbth]
\includegraphics[width=8.5cm]{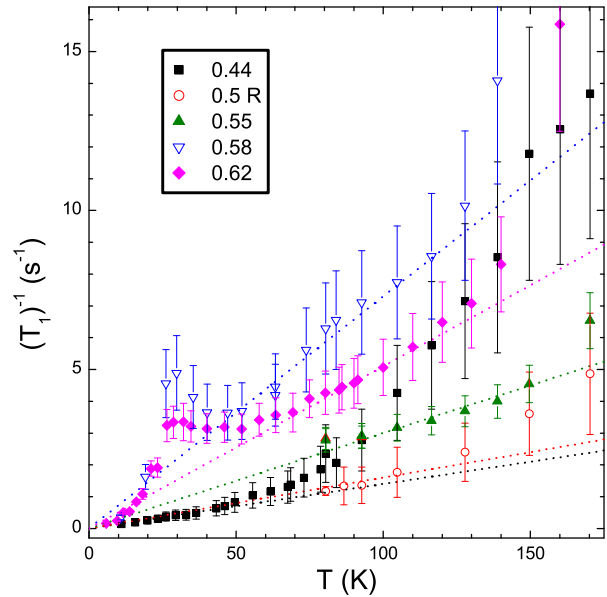}
\caption{(Color online) Temperature dependence of the sodium $T_1^{-1}$ relaxation rate for all studied dopings, along the $c$ direction and at low temperature, with dotted lines crossing the origin as visual guides for the metallic contribution to relaxation.}
\label{fig:T1_lowT}
\end{figure}

If we restrict the data to low temperatures, as was already shown for $x$=0.58 in the inset of Fig.~\ref{fig:T1_058}, we notice (Fig.~\ref{fig:T1_lowT}) that there is a good linear dependence of the relaxation rate versus temperature, clearly establishing the sole contribution of conduction electrons to the relaxation.
Note however that the $x$=0.58 and $x$=0.62 phases feature additionaly a weak relaxation peak around $T$=30~K, which could not be connected to any magnetic transition such as magnetic ordering and remains unexplained.
In order to analyze this data let us first recall that $(1/T_1)_e$ is directly linked to the Co transverse (i.e. perpendicular to the applied field) electronic spin fluctuations through:\cite{Moriya1963}
\[
\left(\frac{1}{T_1T}\right)_e \propto \sum_{\mathbf q} {\left|A_{\perp}({\mathbf q})\right|}^2 \frac{\chi^{''}_{\perp}({\mathbf q},\omega_0)}{\omega_0}
\]
where $A_{\perp}({\mathbf q})$ is the transverse hyperfine form factor and $\chi^{''}_{\perp}({\mathbf q},\omega_0)$ is the transverse dynamic susceptibility at the Larmor pulsation.
In the classical free electron metal, the dynamic susceptibility is $\mathbf q$-independent up to $q=k_F$, and in a case where the hyperfine couplings purely probe on site fluctuations (i.e. $A$ is $\mathbf q$-independent), the product $T_1 T K^2_{spin}$ is constant (Korringa law) and equal to $S_0$=$(\hbar / 4 \pi k_B)(\gamma_e/\gamma_n)^2$, with $\gamma_{e/n}$ the electronic and nuclear gyromagnetic ratio.
In the case where magnetic correlations are important, the Korringa ratio $R$=$S_0 / (T_1 T K^2_{spin})$ may be no more temperature-independent and usually deviates from unity.\cite{Moriya1963}
Indeed, $R$ takes advantage from the fact that the spin shift probes the susceptibility at $\mathbf q$=0, while $T_1$ probes all $\mathbf q$ values.
In the simple case where $A$ is $\mathbf q$-independent, if the fluctuations are predominantly ferromagnetically correlated ($\mathbf q$=0), the shift will be enhanced compared to $T_1^{-1}$, thus $R$ will be lower than 1. On the contrary if the enhancement of the susceptibility occurs at nonferromagnetic wavevectors ($\mathbf q$$\ne$0, such correlations will be named antiferromagnetic, in a general sense, from now on), then the situation is reversed and $R$ becomes higher than 1.

\begin{figure}[pbth]
\includegraphics[width=8.5cm]{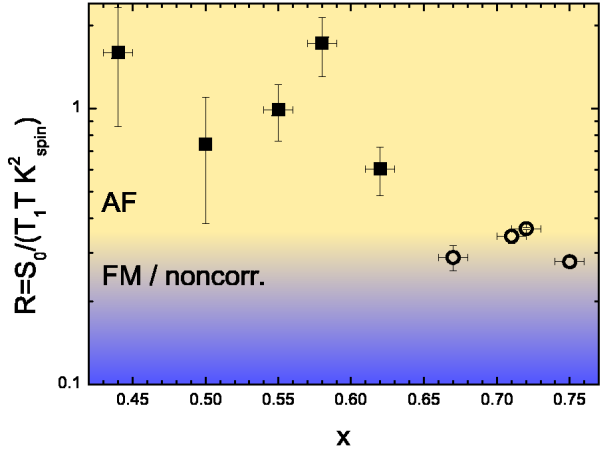}
\caption{(Color online) Doping dependence of the low temperature ($T$$\le$100~K) Korringa ratio $R$ for Na nuclei. Filled squares are from this study, open circles have been extracted from Alloul {\it et al.}.\cite{Alloul2008} The lower region (blue) corresponds to our calculation of the $R$ value expected for noncorrelated or ferromagnetically correlated metals (see text), while the upper region (orange) corresponds to nonferromagnetically correlated metals.}
\label{fig:Korringa}
\end{figure}

Here, on contrary to the situation encountered at higher Na content\cite{Alloul2008},
the sodium shift varies only moderately in the temperature region $T$=50--120~K, or lower for $x$=0.44 (see Fig.~\ref{fig:KNa}).
The $T$-linearity of $T_1^{-1}$ thus makes $R$ fairly temperature-independent in these phases.
Note again that this analysis is restricted to temperature regions below the onset of sodium diffusion, and excluding the weak relaxation peaks around $T$=30~K.
We plot the doping dependence of this low temperature Korringa ratio on Fig.~\ref{fig:Korringa}, with vertical error bars to be understood again as a confidence interval with respect to the fitting of the $T_1$ relaxation measurements.
The $R$ values for $T$=50--100~K and 0.67$\le$$x$$\le$0.75 from Alloul \ETAL\cite{Alloul2008} have been added for comparison.
$R$ is found to be higher than for the Curie-Weiss phases, with no further clear link to $x$.
While this is compatible with a change of correlation regime at $x^*$, a more quantitative assessment is needed, to check whether the ferromagnetic character of in-plane correlations at $x$$\ge$0.67 is lost below $x^*$ and whether new correlations (or a lack thereof) emerge.

Indeed, while comparing $R$ to unity is legitimate in the above case of a free electron metal, we need here to take into account the actual $\mathbf q$ dependence of the Na hyperfine coupling, as $A_{\perp}({\mathbf q})$ may filter strongly certain $\mathbf q$ contributions due to geometrical effects. This is for instance the case in cuprates where planar oxygens filter out antiferromagnetic fluctuations.\cite{Mila1989}.
By calculating $A_i(q) = \sum_{j} A_{ij} \, e^{-{\mathbf q} \cdot {\mathbf r_{ij}}}$ where $j$ describes the Co ions to which the Na site $i$ is coupled, one observes that partial filtering occurs for nonferromagnetic (AF) fluctuations, possibly up to full suppression at certain $q$ points, while ferromagnetic (FM) fluctuations are unfiltered, the exact filtering profile depending on the weighting $A_{ij}$ of the hyperfine paths.
A similar effect is shown by Ning and Imai for the oxygen sites.\cite{Ning2005}
Therefore, it can be assumed that we are severely underestimating the effect of fluctuations at $\mathbf q$$\ne$0 when comparing the measured $R$ ratio to 1.
A better point of comparison is then the expected $R$ value for the $\mathbf q$-independent susceptibility of the noncorrelated metal {\it with the same $A_{\perp}({\mathbf q})$ profile}.
For both kinds of probed Na sites (Na1 and Na2), and depending on the exact weighting of the hyperfine paths, we estimated that this threshold between correlations regimes lies in the range $R_{limit}$=0.15--0.25, following a computation analogous to that of Millis \ETAL\ in cuprates.\cite{Millis1990}
This indicates that AF correlations dominate in the 0.44$\le$$x$$\le$0.62 phases.
For the previously studied $x$$\ge$0.67 phases the temperature dependences of $K$ and 1/$T_1$ do not scale as a Korringa law, and  the correlation between these quantities  clearly establish the occurrence of 2D FM correlations.\cite{Alloul2008}
The proximity of the corresponding data points to $R_{limit}$ could then raise doubt about our calculation of the latter.
However, due to the strong filtering of AF correlations by $A_{\perp}({\mathbf q})$, $R_{limit}$ reflects essentially the $q$=0 contributions to the free electron metal susceptibility.
This can be seen as turning the free electron metal into a FM-correlated metal from the point of view of $(1/T_1)_e$, thus the closeness in values.
So our conclusion on correlations for 0.44$\le$$x$$\le$0.62 stands, and the present study shows that the marked difference in the temperature dependence of the spin susceptibility between the high $x$ phases and those with $x$$<$$x^*$ is associated with a change from dominantly ferromagnetic 2D correlations to dominantly antiferromagnetic correlations.
We cannot conclude directly from the $T_1$ data whether the AF correlations are 2D for $x$$<$$x^*$, but the absence of 3D magnetic order at low temperature, except for $x$=0.5, is a strong indication that the dimensionality of the dominant AF couplings is smaller than or equal to 2.

\begin{figure}[pbth]
\includegraphics[width=8cm]{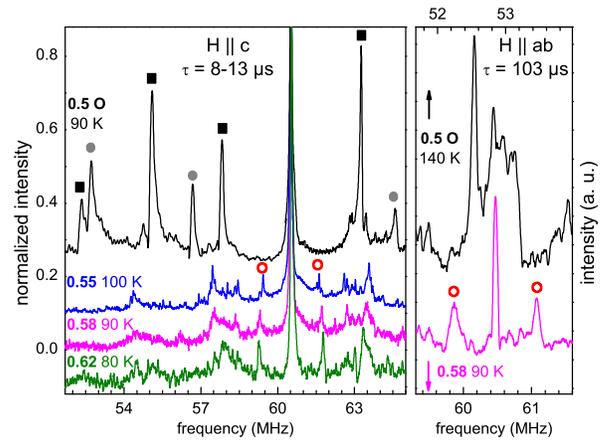}
\caption{(Color online) Cobalt NMR spectra at various dopings with the applied field $H$ along the $c$ axis (left panel) and along the $ab$ plane (right panel).
Filled squares and circles indicate quadrupolar satellites from the A and B sites of the $x$=0.5 phase, while open circles correspond to the new Co sites for $x$$\ge$0.55.
As the $x$=0.5 spectrum has been taken in different experimental conditions, its central line has been brought into coincidence with the others ($\nu_0$=59.3~MHz) by shifting the spectrum, in order to allow for comparison of the quadrupolar splittings.
For $H$ in the $ab$ plane, this has been done so that the spectra coincide at K=2\%, the shift of the newly detected Co sites (see text). Vertical shifts (0.1 along $c$, arbitrary along $ab$) have been applied for clarity.}
\label{fig:specCo}
\end{figure}

\section{Evolution of cobalt charge and spin states}
\label{sec:Co_states}

Another question relative to cobaltates at intermediate dopings is the evolution of cobalt charge and spin states responsible for the observed changes in electronic behavior.
For the $x$=1 and $x$=0 limits, it is known\cite{Lang2005,Vaulx2007} that only one cobalt state is present, with a low-spin $S$=0 3+ state in the first case and a proposed $S$=1/2 4+ state in the second case.
This homogeneous doping situation is retained\cite{Mukhamedshin2005} for $x$$\approx$0.35, both dry and hydrated, suggesting a 3.65+ valence, although the evolution upon hydration remains controversial.\cite{Kubota2004,Takada2004,Milne2004}
Despite suggestions of a possible 3+/4+ charge ordering at low-temperature, we showed in a previous paper\cite{Bobroff2006} that the $x$=0.5 phase has two fairly similar cobalt sites with valence 3.5$\pm \epsilon$ ($\epsilon$$<$0.2).
In contrast, the Curie-Weiss $x$=0.67-0.75 phases feature\cite{Mukhamedshin2005,Alloul2008} several cobalt sites, whose valences have been estimated to be 3+ (nonmagnetic Co) and on average 3.5+ (magnetic Co).
Therefore, it is important to check whether the change in correlation regime coincides or not with a qualitative change in charge distribution within the Co layers, i.e. whether there is some threshold for the appearance of charge disproportionation.

Using cobalt NMR, such a study can be undertaken as the different cobalt sites are probed independently, showing differences in chemical nature and magnetism through orbital and spin shift measurements, as well as differences in charge environments.
Indeed, due to the fact that cobalt has a nuclear spin larger than 1/2 (I(\CoLIX)=7/2), it presents a non-zero electric quadrupolar moment $Q$, thus yielding sensitivity to the local electric field gradient (EFG), a tensor of eigenvalues $V_{xx}/V_{yy}/V_{zz}$.
Treating this term as a first order perturbation of the Zeeman hamiltonian, it is shown to induce a splitting of each resonance line into 2$I$ lines, the central line being unshifted and the quadrupolar satellites appearing on both sides with the same separation:
$$\nu_i = V_{ii} \frac{3 e^2 Q}{2I(2I-1) h}$$
with $i=x,y,z$ and assuming the external field is applied along the principal directions of the EFG tensor.
The quadrupolar hamiltonian is fully defined by the quadrupolar frequency $\nu_Q = \left| \nu_z \right|$ and the quadrupolar asymmetry parameter $\eta = (\nu_x - \nu_y) / \nu_z$.

Experimental spectra with the applied field along the $c$ direction are shown on the left panel of Fig.~\ref{fig:specCo}, with a short $\tau$ delay between radiofrequency pulses of the spin-echo sequence to ensure all sites are detected.
Such extended spectra were obtained from the recombination of individual Fourier transforms.
Let us first recall the situation\cite{Bobroff2006} for $x$=0.5 (here orthorhombic, although the rhombohedral structure shows a similar spectrum): two sites are detected in equal proportions, with well defined quadrupolar satellites having at most some limited superstructure, and with $\nu_Q$$\approx$2.7~MHz for sites A (filled squares on Fig.~\ref{fig:specCo}) and $\nu_Q$$\approx$4.0~MHz for sites B (filled circles).
As this corresponds to two well-defined charge environments, this is in good agreement with the known commensurate structure which features two cristallographic Co sites.\cite{Huang2004b,Zandbergen2004}
Now turning to the 0.55$\le$$x$$\le$0.62 phases, we observe much more complex quadrupolar structures. Such a result fits with the incommensurate orders of the Na layers detected by x-ray diffraction at $T$=300~K, as the occupational modulation and accompanying distortion will naturally result in a series of more or less different Na environments for Co nuclei.
Note that even the spectrum for the $x$=0.55 sample displays such a structure, despite the room-temperature absence of an incommensurate order as seen by x-ray diffraction.
The similarity with the $x$=0.58 spectrum would then indicate that for $x$=0.55 the modulation is fully developped only below room temperature.
More quantitatively, many of the quadrupolar singularities appear to be spread around the positions of the $\nu_Q$=2.7~MHz peaks of the $x$=0.5 phase, while this may also be true but not as clear around the positions of the $\nu_Q$=4.0~MHz peaks. This would suggest that the incommensurability of these phases would simply result in a modulation of the charge environments at $x$=0.5, i.e. that there would not be any significant charge effect within the Co layers themselves.

However, we note the appearance of two new well-defined satellites with $\nu_Q$=1.2~MHz, on each side of the central line (open circles on Fig.~\ref{fig:specCo}), whose spectral weight increases with increasing $x$. This suggests the progressive appearance of a new charge environment, well-defined in spite of the incommensurability, with the corresponding cobalts having a NMR shift $K_c$$\approx$2\%.
Note that the second and third satellites could not be accurately detected for sensitivity reasons.
To further investigate these new cobalt sites, it is more appropriate to use spectra taken with the external field within the $ab$ directions (i.e. within the layers), since it is known from previous work\cite{Mukhamedshin2005,Bobroff2006} that the NMR shift along the $c$ direction is fairly independent of the probed site, while stronger differences between sites show up in the $ab$ directions.
The right panel of Fig.~\ref{fig:specCo} shows such spectra for $x$=0.5 and $x$=0.58, with a large $\tau$ delay between radiofrequency pulses to suppress quickly-relaxing cobalt sites.
For $x$=0.58, the central line features a single peak with $K_{ab}$=1.97(2)\%, with quadrupolar satellites at 0.60(1)~MHz on each side of the central line.
While the central line for $x$=0.5 is not fully suppressed for such long $\tau$ (note that the peak around 52.5~MHz corresponds to an extrinsic impurity), no comparable structure is found for this doping.
It is therefore the new site that we are probing along $ab$ for $x$=0.58, yielding $\nu_Q$=1.20(1)~MHz and $\eta$=0.00(2). On closer examination of the central line along $c$ at $T$=80~K, we find a shoulder at the very similar shift $K_c$=1.94\%. The new cobalt site thus has a quite isotropic shift, almost identical in value to that of the \CoIII\ site\cite{Mukhamedshin2005,Lang2005} for $x$=0.67 and $x$=1, with a long transverse relaxation time implying weak magnetism.
It has then all the characteristics of a \CoIII\ site. Some small temperature dependence of the shift could be observed, but this is easily explained by a transferred hyperfine term.
Note that $\nu_Q$ increases from 1.10(2)~MHz for $x$=0.55 to 1.25(2)~MHz for $x$=0.62, probably due to small valence changes in the various ions or to the difference in structural modulation.
Measured \CoIII\ fractions evolve from 6\% to 11\%, to be compared in the next section to known results at higher $x$ (lower panel of Fig.~\ref{fig:diagram}).
Electroneutrality for a given $x$ value then yields an average valence of 3.43--3.48 for the other cobalt sites of the three phases.
It thus appears that a fairly homogeneous charge state is maintained until $x$=0.5, beyond which nonmagnetic \CoIII\ sites show up in increasing amounts and replace intermediate valence cobalts.

\begin{figure}[pbth]
\includegraphics[width=7cm]{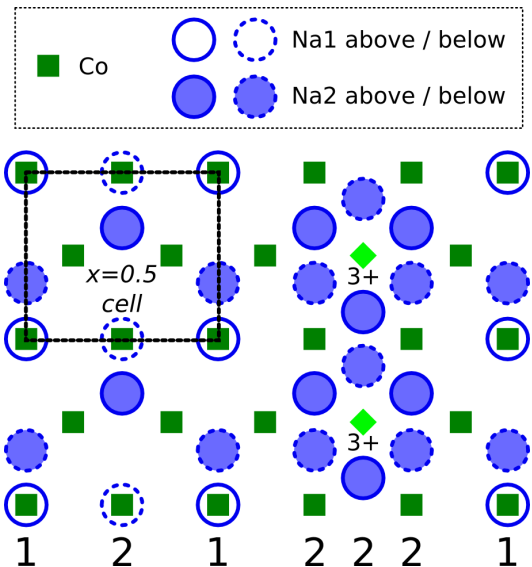}
\caption{(Color online) Proposed structure for $x$$>$0.5, through introduction of stripes of Na2 sites in the $x$=0.5 structure whose unit cell is recalled on the left (bottom labels refer to Na1 and Na2 bands in the upper plane). The limited segregation of three bands of Na2 sites establishes in the middle of the stripe a band of Co sites with fully occupied first neighbor Na2 sites as in Na$_1$CoO$_2$. It is proposed that the structure evolves merely by reduction of the distance between such triple bands with increasing Na concentration.}
\label{fig:struct_Co3}
\end{figure}

To explain the apparition of \CoIII\ sites, it is tempting to refer to the presence of extra Na2 bands as compared to the $x$=0.5 case (see Sec.~\ref{sec:structure}).
Indeed, in the case where the extra Na2 bands are correlated along the $c$ direction, as x-ray diffraction is showing at least for $x$=0.58 and $x$=0.62 for $T$=300~K, it is possible to obtain a situation in which some cobalt sites display a Na environment similar to that in Na$_{1}$CoO$_{2}$, with a saturation of the nearest neighbor Na2 sites.
As shown on Fig.~\ref{fig:struct_Co3}, this requires that the extra Na2 bands segregate at least three by three within the plane.
Note that such a stripe segregation of sodium maintains the existence of quasi-1D $x$=0.5-like domains next to the Na2 stripes (see Fig.~\ref{fig:struct_Co3}). The distortion of the Na lattice in these domains due to the proximity to the Na2 stripes, as well as the direct influence of the latter, would then explain the above-suggested break-up of the $x$=0.5 quadrupolar spectrum into the more complex situation at higher $x$.
Furthermore, as is observed, the \CoIII\ sites would still be probing the magnetism of the more magnetic Co sites neighboring the Na2 stripes, through transferred hyperfine terms. All these arguments support this stripe segregation model, while opposing a scenario where massive segregation of the Na2 bands would occur.

Beyond this qualitative agreement, one may estimate the number of \CoIII\ sites that should arise from such a structure.
Considering commensurate structures featuring one triple-band stripe such as the one on Fig.~\ref{fig:struct_Co3} and $n$ halves of the $x$=0.5 lattice unit cell, it is found that $x=(n+3)/(2n+4)$.
In order to retain at least a full $x$=0.5 lattice unit cell between the Na2 stripes, as discussed above, one obviously needs to set $n$=2 or higher, which happens to correspond to $x$=0.63 or lower (0.60/0.58/0.57...), i.e. values fairly similar to the dopings studied here.
Counting the \CoIII\ sites in the middle of the Na2 stripe then yields a \CoIII\  concentration $y=x-0.5$. This corresponds to a slope half of that in the case of the simple substitution of 3.5+ sites by 3+ sites as $x$ is increased.
This is in very good agreement with the measured fractions, as can be seen from the corresponding visual guide on the lower panel of Fig.~\ref{fig:diagram}.
This triple-band stripe scenario is quite appealing, but it only stems from a simple analysis of the NMR data. We expect that more refined structural studies will allow to establish the validity of this suggested structure.

\section{Phase diagram}
\label{sec:diagram}

\begin{figure}[pbth]
\includegraphics[width=8.5cm]{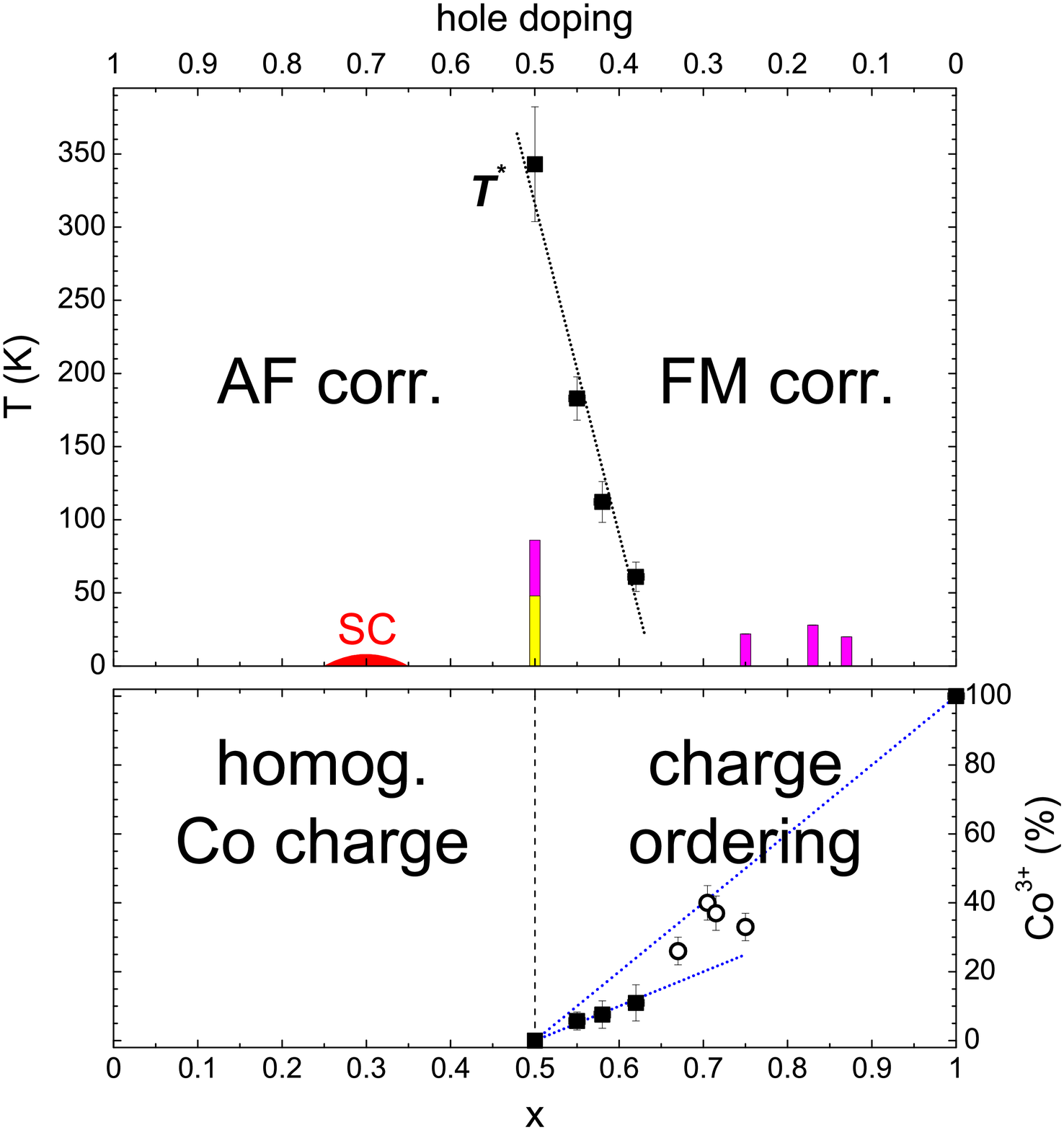}
\caption{(Color online) Proposed phase diagram for the Na cobaltates. In the top panel we report the temperature $T^*$ for which $\chi_{spin}$ is maximal (the dotted line is a visual guide), as well as the nature of spin correlations in the two distinct Na concentration ranges. The known superconducting and magnetically long-range ordered phases are also indicated. In the bottom panel, the concentration of \CoIII\ as determined by NMR is plotted versus $x$, the Co charge being homogeneous up to $x$=0.5 and undergoing disproportionation above (the $x$=1 point and the open circles are from previous studies of our group\cite{Lang2005,Alloul2008}). The dotted lines are visual guides, the high slope ($y=2x-1$) corresponding to the simple substitution of 3.5+ sites by 3+ sites, while the low slope ($y=x-0.5$) corresponds to 3+ sites appearing in the middle of Na2 stripes.}
\label{fig:diagram}
\end{figure}

Our present findings and the previous related results of our group lead us to propose a new phase diagram (Fig.~\ref{fig:diagram}) for sodium cobaltates.

In the upper panel are summarized results relevant to magnetic correlations in the paramagnetic state, as well as the already known superconducting (SC) and long-range magnetically ordered phases.\footnote{For $x$$\approx$0.8--0.9, we have used $x$ determinations not only from measurement of the $c$ parameter but also from that of modulation vectors, in a similar fashion to the $q(r^*)$ measurement for 0.44$\le$$x$$\le$0.62. Details will be published elsewhere.}
Using sodium NMR, we have shown that it is around $x^*$$\approx$0.6, not $x$=0.5, that there is a significant change in the correlation regime, as the uniform spin susceptibility evolves from the well-known Curie-Weiss behavior at high $x$ towards a much less temperature dependent behavior.
For $x$$<$$x^*$, and down to at least $x$=0.5, the susceptibility displays not simply a Pauli behavior as had previously been considered, but increases with increasing $T$ up to a temperature $T^*$ beyond which it decreases.
The energy scale defined by $T^*$ decreases roughly linearly with increasing $x$ and vanishes as $x$$\rightarrow$$x^*$, with the more precise value $x^*$=0.63--0.65.
Note that $x^*$ lies unfortunately in a range of sodium composition for which we could not synthesize single phase samples.
From $T_1$ measurements, we observe metallic behavior for 0.44$\le$$x$$\le$0.62 at temperatures below $T$=50--150~K, i.e. when Na ionic diffusion is non-existent, with modified Korringa ratios of $R$=0.6--1.8, significantly higher than $R$=0.3 measured\cite{Alloul2008} for Curie-Weiss phases at $x$$>$$x^*$.
Taking into account the strong filtering at $\mathbf q$$\ne$0 of the hyperfine form factor, we evidence that for $x$$<$$x^*$ the spin dynamics are not those of a noncorrelated Pauli metal.
The spin fluctuations rather display predominantly antiferromagnetic correlations at low temperature (in the sense of $\mathbf q$$\ne$0), on contrary to the known ferromagnetic character for $x$$>$$x^*$.
Together with results obtained for the superconducting phase Na$_{0.35}$CoO$_2 \cdot y$H$_2$O in its normal state\cite{Ihara2005,Zheng2006a} and for the $x$=0 phase,\cite{Vaulx2007} our results show that $x^*$ marks the low-temperature boundary between a low-$x$ domain with AF correlations and a high-$x$ domain with FM correlations, with the disappearance of any magnetism for $x$=1.
As it appears that a Curie-Weiss behavior may be restored at temperatures high above $T^*$, we suggest on Fig.~\ref{fig:diagram} that $T^*$ could be a correlation crossover temperature.\footnote{One would eventually like to see this crossover in the spin dynamics on the $T_1$ data. However, the actual recovery of a Curie-Weiss-like susceptibility occurs only for $T$ of the order of a few times $T^*$, as can be seen in Fig.~\ref{fig:KNa}. At such temperatures, even for the $x$=0.62 phase which displays the lowest value of $T^*$, the Na nuclear relaxation is totally dominated by the Na ionic diffusion contribution.}

In the lower panel are results relative to the cobalt charge and spin states. While there is as already known an homogeneous cobalt state for $x$$\le$0.5, including in the hydrated superconducting state, we have established that nonmagnetic \CoIII\ states appear for higher sodium content, in line with previous measurements\cite{Mukhamedshin2005,Alloul2008} in Curie-Weiss phases (open circles on the figure).
In the Na2 stripes picture suggested above, we explain the \CoIII\ sites for 0.5$<$$x$$\le$0.62 by the creation of cobalt sites with saturated first-neighbor Na sites, as in the nonmagnetic $x$=1 phase where all cobalts are 3+.
Cobalt sites outside of the Na2 stripes are expected to be fairly similar to cobalt sites in the $x$=0.5 phase, at least from a valence point of view.
This situation corresponds to a {\it de facto} charge ordering, which does not come from an intrinsic charge inhomogeneity of the Co layers.

\section{Discussion}
\label{sec:discussion}

While this new phase diagram hopefully gives a clearer picture of the physics in sodium cobaltates, an important question is the physical origin of the crossover in magnetic correlations.
The existence of two domains has been suggested\cite{Marianetti2007} to result from the direct influence of sodium ions on the electronic states of the Co ions in their immediate vicinity.
In a LDA+DMFT approach, Marianetti and Kotliar show indeed that the cross-over is linked to a change in screening of the Na vacancies potential.
However, as only $x$=1/3 and $x$=3/4 were investigated, it is not clear whether this model would result in two well-delineated regions in the phase diagram.
In view of the rapid rise in temperature-dependence of the susceptibility above $x^*$, and having also measured an upturn of the electronic specific heat coefficient $\gamma$ above $x^*$, Yoshizumi {\it et al.} have proposed\cite{Yoshizumi2007} that a sudden increase of the density of states at the Fermi level occurs.
Such an abrupt increase has been predicted\cite{Korshunov2007} by Korshunov {\it et al.} to occur for a Na content in the 0.56--0.68 range, as a result of the emergence at the Fermi level of a local minimum of the $a_{1g}$ band near the $\Gamma$ point.
Whereas for $x$$<$$x^*$ fluctuations at $\mathbf q$$\ne$0 would be favored, the growth of the electron pocket above $x^*$ would suppress the corresponding Fermi surface instabilities and give rise to FM correlations.
Therefore, in this metallic picture, our results and those of Alloul {\it et al.}\cite{Alloul2008} are well explained, contrary to initial LSDA predictions\cite{Singh2003} which favor FM correlations on the whole $x$=0.3--0.7 range. Note however that the electron pocket near the $\Gamma$ point has not been observed so far by ARPES.\cite{Hasan2004,Yang2004,Yang2005}

The progressive freezing of low-energy excitations at temperatures below $T^*$ could indicate the presence of a pseudogap, as is well-known in underdoped cuprates such as YBa$_2$Cu$_3$O$_{6+x}$.\cite{Alloul1989}
We note that $T^*$ and the cuprate pseudogap temperature both decrease monotonously when doping away from the Mott limit.
This similarity is striking considering that both layered systems also show superconductivity.
However, there is a large difference as the pseudogap in cuprates is evidenced for hole-doping away from the Mott limit, while here the increase in Na content corresponds to electron doping.
Besides, the progressive freezing of low-energy excitations at low temperature should also be seen in the relaxation measurements, as is observed\cite{Berthier1996} in the cuprates where this occurs for a lower temperature than for $\mathbf q$=0.
While our analysis is here again strongly limited by sodium diffusion, especially for high values of $T^*$, such a decrease is not detected experimentally both for $x$=0.58 ($T^*$$\approx$100~K) and $x$=0.62 ($T^*$$\approx$50~K).
We cannot however conclude decisively, as Na does not probe well the fluctuations at $\bf q$$\ne$0, which may substantially reduce any decrease, on top of which would be the unexplained $T_1^{-1}$ peaks around $T$=30~K.
A better check could possibly be $T_1$ measurements on {$^{59}$Co} which would be more sensitive than Na nuclei to the AF spin fluctuations.
Note that even in this case any decrease may be limited, as the decrease observed in the uniform susceptibility, well probed by sodium, is not as pronounced as it is in the case of the cuprates.
Another possibility for this partial freezing of excitations could be low-dimensionality effects, such as in spin chains, spin ladders or frustrated systems.\cite{Imada1998}
Indeed, due to the striped nature of the Na structural order, the effective dimensionality of the cobalt layers might be lower than two, with interconnected cobalt stripes.

Regarding cobalt charge and spins states, our results suggest a purely structural origin to the apparition of nonmagnetic 3+ ions among magnetic $\approx$3.5+ ions.
A correlation between the Na order and the charge disproportionation in the Co layers is known\cite{Mukhamedshin2004} to apply also for $x$=0.67. Further NMR studies are still needed to decide whether this applies as well for $x$$\approx$0.8, although this is presumably so.
In any case, the Na2 stripe scenario can hardly be sustained for $x$$\ge$0.67 for which not even a single unit cell of the $x$=0.5 structure would separate the Na2 stripes. This is compatible with the fact that the \CoIII\ concentration data from previous studies (open circles on the lower panel of Fig.~\ref{fig:diagram}) exceeds that expected from such a model.
This suggests a distinct type of Na ordering for $x$$\ge$0.67, such as the Na di- or tri-vacancy ordering proposed\cite{Roger2007} for instance by Roger \ETAL.
While the Na ordering has no apparent strong impact on the electronic properties in the paramagnetic state below $x^*$, as shown by the very continuous evolution of the susceptibility behavior, this could suggest that the marked change in susceptibility behavior for $x$$>$$x^*$ is of structural origin.
However, in misfit cobaltates where doping is disordered, as in most cuprates, we have evidenced in our group\cite{Bobroff2007} that Pauli-like and Curie-Weiss behaviors at low and high $x$ are maintained.
In the latter case the metallicity is suppressed, which suggests that the Na ordering is essential in maintaining magnetism in correlated metallic bands rather than on localized electronic states.
Concerning long-range ordered states, it appears that the $x$=0.5 phase shows no peculiar behavior of the Co electronic states in the paramagnetic regime with respect to phases with $x$$>$0.5.
So the absence of low-temperature magnetic transitions in the latter cannot be associated with the presence of the Na2 stripes by themselves.
It is then quite likely that it is the commensurability of the Na superstructure which is at play for $x$=0.5, favoring then a scenario of Fermi surface reconstruction and instability with two types of nearly 3.5+ Co ions displaying distinct frozen magnetism.\cite{Bobroff2006,Balicas2005,Zhou2007}
Quite generally, the interplay of the charge and spin degrees of freedom needs further inspection for distinct Na-ordered phases, the relevance of a mixed localized/itinerant approach\cite{Balicas2008,Gao2007} being possible for some phases.

To conclude, let us emphasize that the most important result of this study is the systematic presence of magnetic correlations in the phase diagram. As discussed above, various proposals might explain theoretically this crossover between differently correlated states. However the suggested crossover in temperature at $T^*$ requires further understanding. The instability of the phases at high temperature limits our ability to investigate this crossover, which might eventually be better probed with inelastic neutron scattering experiments.
Beyond questions specific to sodium cobaltates, further insight on these correlations and on the $T^*$ energy scale should help to better relate the physical properties of the layered cobaltate compounds to those of other strongly-correlated systems.

We would like to acknowledge I. Mukhamedshin for fruitful discussions and collaboration on SQUID measurements.
We acknowledge as well D. N. Argyriou, C. Berthier, V. Brouet, S. Hébert, M.-H. Julien, and P. Mendels for helpful discussions.

*Present address: IFW Dresden, P.O. Box 270016, D-01171 Dresden, Germany


\end{document}